\documentstyle[aps]{revtex}
\begin{document}
\title{A Model for the Occurrence of Lock-ins in Holmium in c-axis Magnetic Fields}
\author{M. O. Steinitz$^{*}$, D. A. Pink,$^{*}$ C. P. Adams$^{**\dagger }$ and D. A.
Tindall$^{**}$}
\maketitle

\begin{abstract}

$^{*}$Department of Physics, St. Francis Xavier University, Antigonish, Nova
Scotia, Canada B2G 2W5

$^{**}$Department of Physics, Dalhousie University, Halifax, Nova Scotia,
Canada B3H 3J5

$^{\dagger }$present address: Department of Physics, University of Toronto,
Toronto, Ontario, Canada M5S 1A7

\newpage

%TCIMACRO{
%\TeXButton{doublespace}{\def\baselinestretch{2.0}\small\normalsize}}
%BeginExpansion
\def\baselinestretch{2.0}\small\normalsize%
%EndExpansion

{\bf Abstract.} We introduce a novel interpretation of the sequence of
commensurate lock-ins of the spiral wave vector, $\tau $, observed in the
helimagnetic phase of holmium in the presence of c-axis magnetic fields.
This is the first successful model for the prediction of these lock-ins. We
show that a model which combines a spin-slip model with an assumption
concerning the spin structure along the c-axis in finite size ``domains'',
yields excellent predictions when the energetics of this system with an
applied {\bf c}-axis magnetic field are considered.
\end{abstract}

\newpage 
%TCIMACRO{
%\TeXButton{doublespace}{\def\baselinestretch{2.0}\small\normalsize}}
%BeginExpansion
\def\baselinestretch{2.0}\small\normalsize%
%EndExpansion
{\bf Introduction}

Holmium has been extensively studied for nearly three decades \cite{Koehler
66} \cite{Koehler 67} \cite{Rhyne 67} \cite{Lee 75} \cite{Jayasuriya 85} 
\cite{Gibbs 85} \cite{Bates 88} \cite{Willis 90} \cite{Tindall 93} and many
types of experiments have been performed to probe its magnetic phases and
structures. Despite the wealth of experimental information, no model has yet
been proposed that accounts for the observations that, {\it in applied
magnetic fields}, lock-ins occur at particular (commensurate) wave vectors
and are not apparent in the absence of such fields. Recent work addressing
this question has been presented by Plumer \cite{Plumer 91} and Jensen \cite
{Jensen 96}. To account for these observations, we propose a model which
combines the spin-slip model \cite{Koehler 66} \cite{Bohr 86}, involving
layers with magnetic orientation organized only into ``singlets'' and
``doublets'' (see below), with a key assumption about the magnetic structure
of the commensurate phases and which builds upon earlier work of Steinitz et
al. {}\cite{Steinitz 95}. A consequence of these ideas is that finite size
effects in the magnetic structure are essential to an understanding of the
lock-ins. In this paper we refer to the coherent structures, the (finite)
size of which defines the wave vectors accessible to the system, as
``domains''. We find that when the energetics of the problem are considered,
we are able to predict the experimentally observed lock-ins in the presence
of applied magnetic fields along the {\bf c}-axis and can account for the
fact that the lock-ins become more pronounced as the field increases, and
are not apparent in zero field.

Much work has been done on a general approach to understanding structures in
systems with competing interactions in terms of (domain) wall energies and
wall-wall interactions \cite{Fisher 87} \cite{Selke 92}. In addition, there
have been recent studies on models of rare-earth systems \cite{Sasaki 92} 
\cite{Seno 94}. None of these approaches, however, have considered
finite-size ``domain'' structures as being of possible significance in
understanding lock-in effects in holmium in the presence of applied magnetic
fields. Indeed, work on rare-earth models has been confined either to zero
field or to a field in the basal plane\cite{Harris 91}. Our model is in the
spirit of the work of Fisher and Szpilka \cite{Fisher 87}. It relates
energies directly to wave vectors, which, because of the proposed
finite-size ``domains'' to be described below, are isolated in k-space.

Holmium has a hexagonal close-packed crystal structure (hcp). Below its
N\'{e}el temperature (T$_{{\rm N}}$ = 132 K) it has a helimagnetic or spiral
magnetic structure down to the Curie temperature (T$_{{\rm C}}$ = 20 K), at
which a {\bf c}-axis component of the moment develops and the wave vector, $%
\tau $, locks in at a value of 1/6, i.e. one turn in 6 unit cells or twelve
layers. (We will always refer to $\tau $ in reciprocal lattice units along 
{\bf c}*, the hexagonal axis.) In the helimagnetic phase, the moments in the
basal planes are ferromagnetically aligned, with a temperature dependent
angle between the magnetic orientations of adjacent planes, leaving the
material antiferromagnetic overall. Generally, this structure appears to be
incommensurate with the crystal structure along the {\bf c}-axis. In zero
field it is found that $\tau $ varies, apparently continuously, from a value
of 1/6 at T$_{{\rm c}}$ to a value which we have found to be 5/18 just at T$%
_{{\rm N}}$ \cite{Tindall 93}.

In the presence of an applied magnetic field we have found that $\tau $ does
not vary smoothly as the temperature is changed, but locks in at certain
commensurate values, i.e. it tends to stay at certain values for a finite
temperature interval before continuing its apparently smooth variation. In a
sequence of papers \cite{Tindall 93} \cite{Noakes 90} \cite{Tindall 932} 
\cite{Tindall 92} \cite{Tindall 94} \cite{Tindall 942}, lock-ins were
reported at 1/5, 2/9, 1/4, and 5/18 in addition to the lock-in at 1/6 at T$_{%
{\rm C}}$ which occurs even in zero field, and one found by Gibbs {\it et
al.\/} \cite{Gibbs 85} at 2/11. A comparison of the field and temperature
values where lock-ins occur with the phase boundaries on a phase diagram\cite
{Willis 90} obtained from magnetization measurements, shows a clear
correspondence. We have also reported the observation of anomalous
``noise'', presumably due to fluctuations, in dilatometric measurements \cite
{Steinitz 92}. This noise disappears at T$_{{\rm N}}$ and at the $\tau $ =
1/4 lock-in in a 3 T {\bf c}-axis field, leading us to associate it with
fluctuations in the spiral turn angle of coherent domains (given the
correspondence between thermal expansion and the variation of $\tau $).

\smallskip

{\bf Theory }

{\bf 1. Localized-spin Hamiltonian and the Spin-slip Model}

\smallskip A model based on spin-slips has previously been used with great
success to describe the observed Fourier components of the magnetic
structures of holmium \cite{Bohr 86} \cite{Cowley 91} \cite{Jensen 90}. The
spin-slip model arose from the observation that at the Curie point, where $%
\tau $ = 1/6, there is not, in fact, a constant turn angle between the
planes. Instead the magnetization vectors of the planes are bunched up or
paired into doublets which lie near the six easy magnetization directions 
\cite{Koehler 66} (see Figure 4 of McMorrow et al., 1993 for a clear picture
of such an arrangement) \cite{McMorrow 93}. As the temperature increases,
some of the doublets break up into singlets, or unpaired planes. Each
singlet is referred to as a spin-slip. By using various combinations of such
spin-slips one can find a spin-slip structure that matches any $\tau $ value
that is a rational fraction. If a doublet is represented by a ``2'' and a
singlet by a ``1'' then, for instance, 222221 gives a value of $\tau $ =
2/11. (Note that there are two basal plane layers per unit cell along {\bf c}%
, so that 1 turn in 11 layers makes 2 turns in 11 cells.) This choice of
structure is not unique and there can be several spin-slip structures which
yield the same $\tau $.

In previous work \cite{Steinitz 95} it was observed that, upon considering
the sequence of rational numbers representing commensurate $\tau $'s, up to
some maximum denominator (dictated in real materials by, for example, a
finite ``domain size'' or correlation length), certain numbers are
relatively isolated from their neighbors, i.e. they are the only occupants
of gaps in the sequence. This is shown schematically in Figure 1, where we
have drawn a vertical line at each rational fraction up to an arbitrarily
chosen maximum denominator of 60 chosen for an acceptable density for
printing purposes. As was pointed out\cite{Steinitz 95} , a number of about
150 corresponds to a reasonable estimate of domain size or coherence length
from estimates of distances between imperfections such as impurities or
dislocations which might act as pinning sites. It is well-known that the
gaps between the fractions close only very slowly as the maximum denominator
increases. This pattern arises from the fact that if the rational fractions
considered are limited to those with some maximum denominator, the sequence
is a Farey series \cite{Hardy 38} in which the interval between sequential
elements with denominators {\it a\/} and {\it b\/} is known to be 1/{\it ab\/%
}. Thus the most isolated values are those with small denominators and
correspond closely to the observed locked-in values of $\tau $.

This observation does not, however, give any indication as to why some of
these fractions with low denominators correspond to observed lock-ins and
why some do not. The principal contribution of this work is to propose an
answer to this by calculating the energies of spins in zero and non-zero 
{\bf c}-axis fields. The calculation of these energies will be based upon
the spin-slip model in the temperature regime of interest. In order, then,
to develop our model of magnetic-field induced lock-ins, we must establish
the validity of the spin-slip model, which we do below, using computer
simulation.

The simplest model for localized spins interacting via a Heisenberg exchange
interaction is\cite{Elliott 61} :

\begin{eqnarray}
H &=&\frac{-J_0}2\sum_{i,\delta }{\bf S}_{ni}\cdot {\bf S}_{n,i+\delta }-%
\frac{J_1}2\sum_{i,\rho }{\bf S}_{ni}\cdot {\bf S}_{n+1,i+\rho }-\frac{J_2}2%
\sum_{i,\zeta }{\bf S}_{ni}\cdot {\bf S}_{n+2,i+\zeta }  \nonumber \\
&&+\Delta \sum_{n,i}({\bf S}_{n,i}^z)^2+2K_6\sum_{n,i}(\cos (6\phi _{{\rm ni}%
})+1)-H\sum_{n,i}{\bf S}_{n,i}^z
\end{eqnarray}
where S$_{{\rm ni}}$ is the spin vector at the i-th site of the n-th layer
in the basal plane, with the direction of the spin vector given by (\ $%
\theta _{{\rm ni}}$, $\phi $$_{{\rm ni}}$). {\it J}$_{{\rm 0}}$ is the
ferromagnetic exchange interaction between nearest neighbour spins in the
basal plane, {\it J}$_{{\rm 1}}$ and {\it J}$_{{\rm 2}}$ are the
interactions between nearest and second nearest neighbor spins on layers n$_{%
{\rm m1}}$ and n$_{{\rm m2}}$ respectively. $\Delta $ and K$_{{\rm 6}}$
represent crystal field parameters for {\bf c}-axis and six-fold basal plane
anisotropy. In order to verify the validity of the spin-slip model, we
carried out Monte Carlo (MC) simulations of classical spin ordering in this
system over a wide temperature range using the Metropolis algorithm\cite
{Binder 84}.

We considered a simple hexagonal lattice with the spin system described by
the hamiltonian (1). We used periodic boundary conditions in the {\bf a-b}
plane, which was represented by either a 10 x 10 or 15 x 15 lattice, and
open boundary conditions along the {\bf c}-axis, where the number of layers
possessing six-fold symmetry, {\it n}, ranged from 100 to 360. Both the 10 x
10 and the 15 x 15 lattices gave the same results. Spins were visited in a
sequence which chose the layer and the site within the layer using a
pseudo-random number generator. We chose spin magnitude = 1.0, with $J_0=1.0$%
, $J_1=1.0$, $J_2=-1.5$ in order to yield a ``classical'' turn angle of $%
\phi =\pi /5$. We chose $\Delta =0.2$ simply in order to drive the spins out
of the basal plane with an external field, $H$, not too large, and $K_0=0.5$
so that the depth of the six-fold anisotropy was comparable to the exchange
energies. We initialized the system for 1000 MC steps and performed averages
over 5000 MC steps. For values of n$\geq 140$ we found no difference in the
results. The small size of the basal-plane lattice had no effect:
fluctuations in spin direction within a given basal-plane were $\sim 2^o$.
MC runs greater than 5000 MC steps yielded no change in the results.

Figure 2 shows typical results for the instantaneous turn angles per layer,
averaged over all turn angles within a layer, for a non-zero field, H,
applied along the {\bf c}-axis. It can be seen that the distribution is
composed predominantly of ``singlets'' and ``doublets'', though we did not
carry out an analysis to confirm this. It is easily confirmed that the
average turn angle, averaged over all layers, is equal to the turn angle per
layer, calculated analytically on the assumption that it is a constant from
layer to layer. This last observation is correct within the confidence
limits of the simulation.

\smallskip

{\bf 2. A Model of Magnetic-Field-Induced Lock-ins.}

In our model, we assumed that a spiral or conical arrangement of classical
spins exists which is composed entirely of ``systems'' of ``singlets'' and
``doublets'', labelled as 1 and 2 respectively, in which there are n$_{{\rm %
kl}}$ systems of type k immediately adjacent, along the positive {\bf c}%
-axis, to a system of type l (k,l = 1,2). Thus the number of doublets with a
singlet adjacent to it along the {\bf c}-axis is n$_{{\rm 21}}$ We assume,
initially, that all spins are localized along the directions of the local
minima of the six-fold basal-plane anisotropy energy. With this assumption.
when the system is of infinite extent along the {\bf c}-axis, then the
following relations hold:

\begin{equation}
n_{21}=n_{12}\text{ , }n_{12}+n_{22}=n_2\text{ , }n_{11}+n_{12}=n_1 
\eqnum{2}
\end{equation}
where {\it n}$_{{\rm 1}}$ and {\it n}$_{{\rm 2}}$ are the total numbers of
singlets and doublets respectively. We note that the total number of layers, 
{\it m}, the total number of turns, {\it n}, and the average turn angle per
layer, $\phi $ are given by

\begin{equation}
m=n_1+2n_2\text{ , }n=\frac{n_1+n_2}6\text{ , }\phi =2\pi n/m  \eqnum{3}
\end{equation}

The third of these equations follows from the assumption that in the absence
of an applied magnetic field, all spins are located at the local minima of
the six-fold basal plane anisotropy energy. From our simulations, however,
we know this to be false: the doublets are ``split'' slightly, while the
singlets can be located away from the local minimum. This arises because of
the exchange couplings. We shall use this observation in what follows:

We assumed that a doublet is split slightly and calculated the dependence of
the energy of a doublet, in the six-fold basal plane anisotropy field (the
term in $K_6$ on the right hand side of equation (1)) only, as a function of
the tilt angle $\theta $\ , away from the {\bf c}-axis, in a magnetic field,
H, applied along the {\bf c}-axis. We assumed that each spin in the doublet
possessed the same value of \ $\theta $. When $H=0$, $\theta =\pi /2$. We
assumed that the doublet splitting and orientation projected onto the basal
plane did not change with \ $\theta $ and found that this energy $\sim
1/2\sin ^2\theta $ \ near the local minima of the potential. A similar
result holds for singlets, if they do not lie precisely at one of the minima
of the six-fold anisotropy energy and if they do not change their
orientation, projected onto the basal plane, with $\theta $. If the singlets
and doublets lie precisely at the minima of the six-fold basal-plane
anisotropy, then there will be no $\theta $ dependence of the energy as the
external field is turned on.

All these assumptions are relatively minor. In the case of the assumption
concerning the constancy of the doublet splitting irrespective of the value
of \ $\theta $, we note that the energy will still increase, because a
change in splitting, to reduce the six-fold basal plane anisotropy energy,
will be balanced by increases in the exchange energies.

Our last assumption, however, is the key to our model: {\bf we assume that,
for a given average turn angle, }$\phi ${\bf , the system, composed entirely
of singlets and doublets, will minimize its energy by having singlets on the
two layers which terminate a complete spiral of }$2\pi k${\bf \ radians,
where }$k${\bf \ is the smallest integer which can achieve this, and that
these two singlets will be located at a local minimum of the six-fold basal
plane anisotropy field}. Accordingly, when a field, H, is applied, those two
singlets do not increase their six-fold basal plane anisotropy energy. This
assumption is shown schematically in Figure 3 for m=8, n=1, $n_1$ = 4 and $%
n_2$ = 2, making the (112112) structure. Note that a doublet could not be at
an end of a sequence because the two spins are ``split'' nearly, or exactly,
symmetrically and so will not then lie in a minimum of the basal plane
anisotropy. The figure shows one instantaneous configuration, as appears in
the simulations (of which an example is shown in Figure 2), where
interactions cause the local spins to lie away from the minima of the
hexagonal anisotropy field.

In order to see whether such a spin arrangement is stable under a rotation
of all the spins of a complete spiral, let us consider the arrangement shown
in Figure 3. We shall consider only the energy of the six-fold basal plane
anisotropy field. We assume, for simplicity, that every member of a doublet
lies at angles $\phi _2$ on either side of the local minimum, and that half
of the singlets are offset from the local minimum by $\phi _1$and the other
half by -$\phi _1$. This last comment does not apply to the two spins at
either end which lie at the minimum of the anisotropy field. It is then
trivial to show that the system is indeed stable under an infinitesimal
rotation of all vectors by $\phi $. This provides some justification for our
proposal that such an arrangement is set up in antiferromagnetic holmium.

\newpage With this assumption and the results preceding it, the energy per
layer, in units of {\it J}$_{{\rm 1}}${\it S}$^{{\rm 2}}$, becomes:

\begin{eqnarray}
\frac E{mJ_1S^2} &=&-\frac{n_2}m-\frac{6n}m(\cos \phi +\cos ^2\phi -\cos
\phi \cos ^2\theta )+\frac{n_1}m\frac{J_2}{J_1}  \eqnum{4} \\
&&-\frac{J_2}{J_1}(\frac{12n}m\cos ^2\theta +\frac{2n_2}m\cos \phi +\frac{%
2n_1}m\cos ^2\phi -\frac{2n_2}m\cos \phi \cos ^2\theta  \nonumber \\
&&-\frac{2n_1}m\cos ^2\phi \cos ^2\theta )+\frac \Delta {J_1}\cos ^2\theta -%
\frac H{J_1S}\cos \theta  \nonumber \\
&&+\frac{K_6}{J_1S^2\sin ^2\theta }(\frac{n_1-1}m+\frac{n_2}m)  \nonumber
\end{eqnarray}
where S is the spin magnitude and we have omitted the term in {\it J}$_0$,
since this does not change with {\it H}. The term in $n_1-1$ allows for the
two singlets at each ``end'' of a complete spiral of total turn angle $2\pi
k $.

If we minimize $E/mJ_1S^2$ with respect to the turn-angle, $\phi $, we
obtain:

\begin{equation}
\frac{J_2}{J_1}=\frac{-3n/m}{\frac{n_2}m+\frac{2n_1}m\cos \phi }  \eqnum{5}
\end{equation}

If we minimize $E/mJ_1S^2$ with respect to the tilt-angle, $\theta $, away
from the {\bf c}-axis, then we obtain a relationship between {\it H} and $%
\theta $,

\begin{eqnarray}
\frac H{J_1S} &=&(\frac{2K_6}{J_1S^2\sin ^4\theta }(\frac{n_1-1}m+\frac{n_2}m%
)+\frac{2\Delta }{J_1}-\frac{12n}m(1-\cos \phi )  \eqnum{6} \\
&&-\frac{2J_2}{J_1}(\frac{12n}m-\frac{2n_2}m\cos \phi -\frac{2n_1}m\cos
^2\phi ))\cos \theta  \nonumber
\end{eqnarray}

We then substitute for H in (4) to give,

\begin{eqnarray}
\frac E{mJ_1S^2} &=&-\frac{n_2}m-\frac{6n}m(\cos \phi -\cos ^2\theta +\cos
\phi \cos ^2\theta )+\frac{n_1}m\frac{J_2}{J_1}  \eqnum{7} \\
&&-\frac{J_2}{J_1}(-\frac{12n}m\cos ^2\theta +\frac{2n_2}m\cos \phi +\frac{%
2n_1}m\cos ^2\phi +\frac{2n_2}m\cos \phi \cos ^2\theta  \nonumber \\
&&+\frac{2n_1}m\cos ^2\phi \cos ^2\theta )+\frac \Delta {J_1}\cos ^2\theta 
\nonumber \\
&&+\frac{K_6}{J_1S^2\sin ^4\theta }(\frac{n_1-1}m+\frac{n_2}m)(\sin ^2\phi
-2\cos ^2\theta )  \nonumber
\end{eqnarray}

\smallskip

{\bf Results}

We have plotted this energy, $\frac E{mJ_1S^2}$, in Figure 4, as a function
of $\phi $ (or $\tau $) for two values of H. It should be noted that \ H
depends implicitly upon $\phi $ so that {\bf spin arrangements with
different turn angles have different arrangements of angles with the c-axis
in non-zero H}. We varied $\phi $ by changing the ratio $\frac{J_2}{J_1}$ .
Experimentally it is known that $\phi $ varies with temperature, implying
that $\frac{J_2}{J_1}$ varies with temperature. In Figure 4 it can be seen
that certain wave vectors having small values of m exhibit local minima,
which are shallow in zero field but which become deeper as a {\bf c}-axis
field is applied. Accordingly, the sheets of ferromagnetically-aligned spins
in the basal planes should exhibit some cooperative effects characteristic
of ``spin-reorientation,'' and lock-in at those wave vectors with local
energy minima. The extent of fluctuations, as the lock-in wave vector is
approached, and the stability of the locked-in state itself, will be
determined by the depth of the local energy minimum compared to energies of
neighbouring wave vectors as well as the density of states as a function of
wave vector. The density of states is suggested by Figure 1, where it is
seen that some wave vectors are relatively ``isolated''. It should be noted
that, in particular, those wave vectors at $\tau $ = 2/11, 1/5, 2/9 and 1/4
are the only ones that are isolated in this way, and that these correspond
precisely to the set of lock-ins observed in neutron diffraction
measurements.

A consequence of this is that in zero magnetic field, measurements will give
the appearance of continuously varying incommensurate $\tau $ values within
the experimental resolution and within our ability to resolve separate peaks 
\cite{Steinitz 95}. Based on our ability to detect the presence of a second,
coexisting phase \cite{Tindall 942}, we can say that any fluctuation or
noise in $\tau $ due to multiphase behaviour is certainly less than 0.002
c*. This is sufficient to hide lock-in effects, at least for higher
temperatures. However, the 2/11 lock-in has been observed even in zero field
below 30 K, although this may be due to the effect of local strain mimicking
the effect of an applied field in this material, which has an immense
spontaneous magnetostriction {}\cite{Rhyne 67}\cite{Plumer 91}.

The wave vector 5/18 is marked in Figure 4 because experimentally, this is
always the ultimate value of the wave vector at T$_N$ in zero field and in
any applied {\bf c}-axis {\it or} {\bf b}-axis field tried to date. This is
a great mystery to us, and we have no explanation whatsoever for it. There
is a significantly wider lock-in in a {\bf b}-axis field \cite{Tindall 92}(7
or 8 Kelvins wide) than any oberserved in {\bf c}-axis fields. The absence
of any lock-in at 5/18 in our work on {\bf c}-axis effects presented here is
in accord with experiment.

\smallskip

{\bf Conclusions}:

The simple model presented here predicts exactly all of the wave vectors of
all the {\bf c}-axis magnetic-field-induced lock-ins observed experimentally
in the hcp helimagnet, holmium. The relative probabilities of observing
lock-ins can be estimated by assuming that there is some characteristic
length, L, along the {\bf c}-axis into which the lattice distances over
which the complete spiral of 2$\pi k$ radians (see above), terminated at
each end by singlets, must ``fit''. Clearly, the number of such 2$\pi k$%
-radian ``units'' will be proportional to 1/m. These results should be
relevant to many other hexagonal antiferromagnets which exhibit helical spin
structures \cite{Diep 94}. This model assumes (a) that the magnetic
structure is sufficiently well-described, for the temperature range of
interest, by ''singlet-doublet'' spin-slip structures, and we have offered
the evidence of computer simulations to support this; and (b) {\bf we assume
that, for a given average turn angle, $\phi $, the system, composed entirely
of singlets and doublets, will minimize its energy by having singlets on the
two layers which terminate a complete spiral of 2}$\pi k${\bf \ radians,
where k is the smallest integer which can achieve this, and that those two
singlets will be located at a local minimum of the six-fold basal plane
anisotropy}. As the wave vector changes, lock-ins will occur at those wave
vectors for which the energy exhibits a sufficiently deep local minimum.
Fluctuations should be observed for wave vectors in the neighbourhoods of
those wave vectors for which lock-in behaviour is manifested at the
temperatures of interest. The wave vectors can be swept by changing the
temperature in a constant {\bf c}-axis magnetic field. Analogous lock-ins
should be observed if the temperature is fixed and the field is swept, so as
to sample those $\tau $ values which exhibit sufficiently-low local values
of energy. The phase boundaries found in magnetization work, however, are
quite steep, making some of these field sweeps quite difficult to carry out.
We note that the energy term in equation (1), which we have described as a
6-fold basal plane anisotropy field, might be enhanced significantly,
because of the large magneto-elastic coupling in these materials, as the
crystal is distorted by the application of a {\bf c}-axis magnetic
field.\medskip

We are pleased to acknowledge the financial support of the Natural Sciences
and Engineering Research Council of Canada and the Canadian Institute for
Neutron Scattering. It is a pleasure to thank Ms. Bonnie Quinn for writing
the code to simulate magnetic ordering in holmium, Mr. Brian Segal for
preparing the figures, and the staff at the Chalk River Laboratories of
AECL, especially Dr. T. M. Holden, for their invaluable assistance.

\newpage {\bf Figure Captions}

Figure 1. The Farey series with denominator up to 60. A vertical line is
drawn at each rational fraction with denominator less than 60 in order to
illustrate the gaps between the fractions. The gaps fill in very slowly as
the maximum denominator increases.\bigskip

Figure 2. Average angle of spin orientation projected onto the n-th basal
plane, $<\phi _n>$, as a function of plane, n, obtained from a Monte-Carlo
simulation of equation (1). The ratio of $J_2/J_1$ was selected to give an
average turn angle of $\pi /5$ per plane as an example. Each basal plane was
represented by a (10 x 10) triangular lattice with periodic boundary
conditions and this simulation used 360 planes with open boundary
conditions, so that correlations due to {\bf c}-axis periodicity were not
introduced. Here, a {\bf c}-axis magnetic field was applied and gave values
of $\theta _n$, the angle measured from the {\bf c}-axis, lying between
about 61 and 66 degrees. The graph shows planes 0 to 140 so that the effect
of the open boundary condition can be seen.

\smallskip

\smallskip

Figure 3. An example of a spin-slip structure (112112) with an average turn
angle of $\pi /4$ per layer ($\tau =\frac 14$) with the spin vectors
projected onto the basal plane. Here, m = 8, n=1, $n_1$ = 4 and $n_2$ = 2.
Note that m is the number of spacings between the two singlets (shown as
heavy lines because they have the same orientation) which are at either end
of the spin structure. Accordingly, one of those end spins is not included
in $n_1$. The dashed lines indicate the local minima of the six-fold basal
plane anisotropy field.\bigskip

Figure 4. The energy per layer, $\frac E{mJ_1S^2}$ , calculated from
equation (7) as a function of the wave vector, $\tau $, in a magnetic field
and in zero field for comparison. The parameters used are $\frac \Delta {J_1}
$ = 2, $\frac{K_6}{J_1S^2}$= 2, and $\frac H{J_1S}$= 0 and $\frac H{J_1S}$ =
20. The Farey series of order 150 was used to generate the wave vectors.
Points with denominators, m 
%TCIMACRO{\TEXTsymbol{<} }
%BeginExpansion
\mbox{$<$}%
%EndExpansion
20 are circled. The important points to note are the isolation of the
low-denominator points (in particular those at 2/11, 1/5, 2/9, and 1/4), the
field dependence of the ``lock-in'' energy, and the difference between the
energy of the isolated low-denominator points and that of the adjacent wave
vectors. The points for H = 20 have been displaced upwards by 10 for display
purposes.

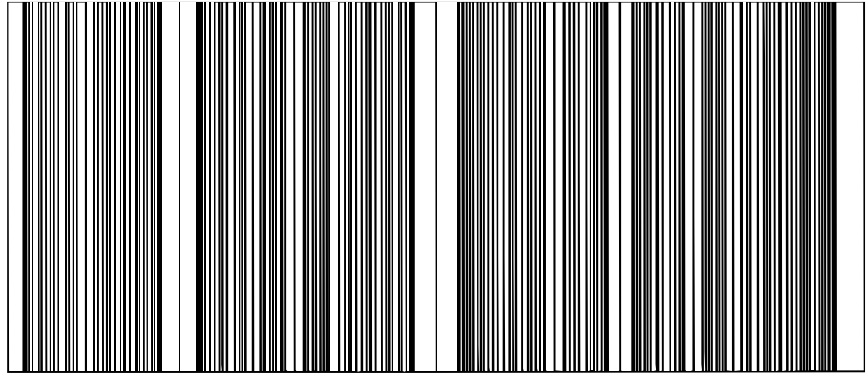
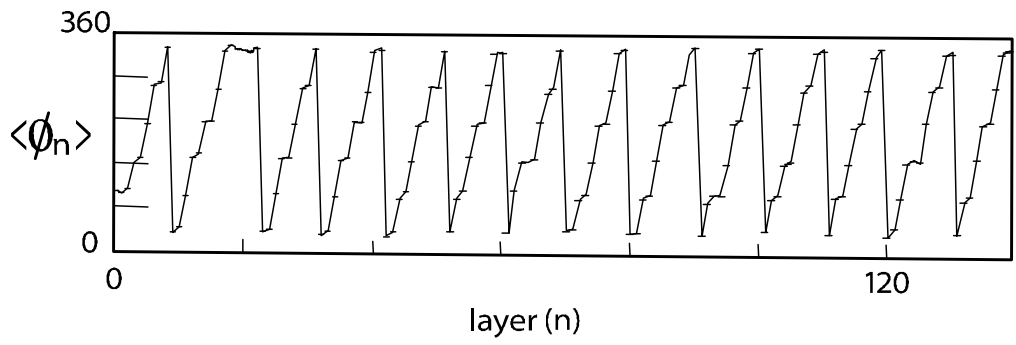
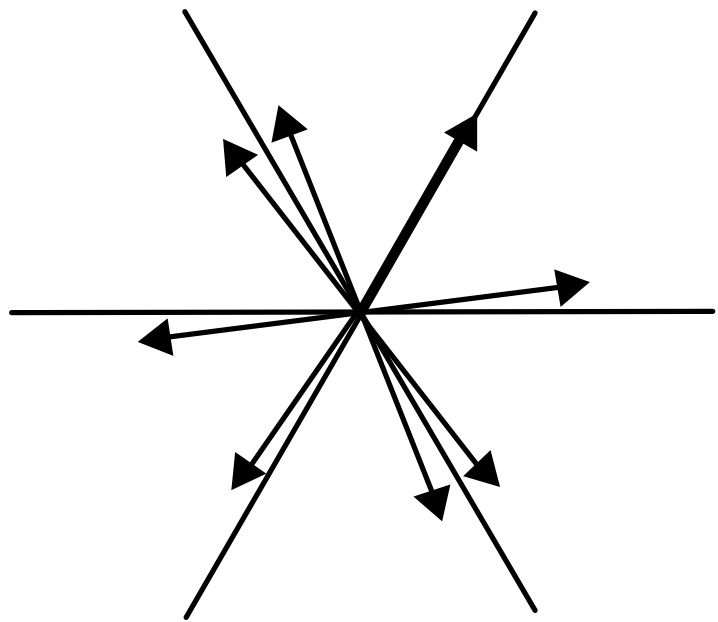
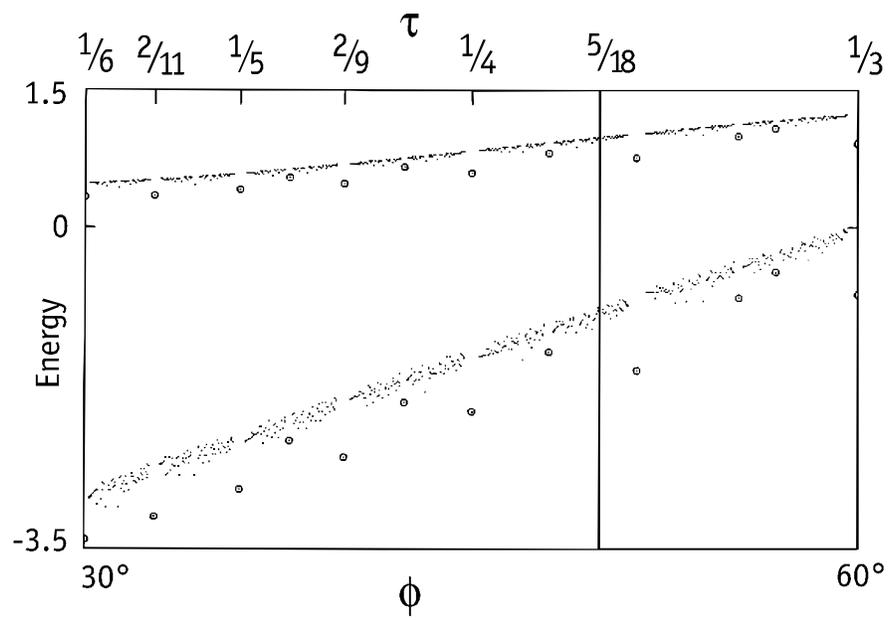
\end{document}